\pdfoutput=1
\documentclass[oupdraft]{bio}
\usepackage{url}
\usepackage{amsmath}
\usepackage{graphicx,psfrag,epsf}
\usepackage{enumerate}
\usepackage{natbib}
\usepackage{url} 
\usepackage{bm}
\usepackage[utf8]{inputenc}
\usepackage[english]{babel}
\usepackage{amsthm}
\usepackage{url}
\usepackage{amsfonts}
\usepackage{multirow}
\usepackage{xcolor}
\usepackage{booktabs}

\usepackage{multirow}
\usepackage{comment}

\begin{document}
\title{Dynamic Models Augmented by Hierarchical Data: An Application Of Estimating HIV Epidemics At Sub-National Level}

\author{Le Bao$^\ast$ and Xiaoyue Niu\\[4pt]
\textit{Department of Statistics, Penn State University, University Park, PA, USA}\\[2pt]
Tim Brown\\[4pt]
\textit{Research Program, East-West Center, Honolulu, HI USA}\\[2pt]
Jeffrey W. Imai-Eaton\\[4pt]
\textit{Center for Communicable Disease Dynamics, Department of Epidemiology, Harvard T.H. Chan School of Public Health, Boston, MA, USA}\\
\textit{MRC Centre for Global Infectious Disease Analysis, School of Public Health, Imperial College London, London, UK}}

\markboth%
{L.Bao et. al.}
{Dynamic Models Augmented by Hierarchical Data}
\maketitle

\footnotetext{To whom correspondence should be addressed. lebao@psu.edu}

\begin{abstract}
{Dynamic models have been successfully used in producing estimates of HIV epidemics at the national level due to their epidemiological nature and their ability to estimate prevalence, incidence, and mortality rates simultaneously. Recently, HIV interventions and policies have required more information at sub-national levels to support local planning, decision making and resource allocation. Unfortunately, many areas lack sufficient data for deriving stable and reliable results, and this is a critical technical barrier to more stratified estimates. One solution is to borrow information from other areas within the same country. However, directly assuming hierarchical structures within the HIV dynamic models is complicated and computationally time-consuming. In this paper, we propose a simple and innovative way to incorporate hierarchical information into the dynamical systems by using auxiliary data. The proposed method efficiently uses information from multiple areas within each country without increasing the computational burden. As a result, the new model improves predictive ability and uncertainty assessment.}
{Hierarchical model; Dynamical systems; HIV epidemics}
\end{abstract}

\section{Introduction}
\label{sect-Introduction}
Recent estimates suggest that great progress has been made in combating the HIV/AIDS epidemic, including a 38\% decline in new infections globally since 2010 and a 52\% drop in AIDS-related deaths since 2010 \citep{UNAIDS2023}. However, it is perceived that among these successes, 
sub-national geographic regions with greatly elevated burden have not benefited equally \citep{UNAIDS2014,UNAIDS2020b}. 
Within high burden countries, geographic variations by province and district can be large, e.g., in Zambia, regional HIV prevalence varied by a factor of three from 6.4\% to 18.2\% in 2013 \citep{ZambiaDHS2015}.
To quantify and efficiently prioritize resource allocations and acknowledge the heterogeneity, we need more accurate information about HIV at sub-national levels. 

Currently, the sub-national estimations are done by applying the epidemic models \textit{independently} using only data from within the area \citep{Mahy2014}. For groups with sparse data, the model sometimes produces inaccurate results with large uncertainty bounds \citep{Lyerla2008,Calleja2010}. One solution is to assume that the parameters of the epidemic models are correlated among areas in a hierarchical framework, thereby efficiently borrowing information from other areas with similar epidemics. However, the epidemic model consists of differential equations that do not have analytic solutions and estimating parameters in the HIV epidemic models is already time-consuming \citep{Raftery2010}. Fitting multiple such dynamical systems with a hierarchical structure would substantially increase the computational cost. To facilitate the joint estimation of multiple sub-epidemics, \cite{bao2023estimating} introduces the dependence of the parameters across multiple dynamical systems and approximates their joint distribution by adjusting the importance sampling weights. However, when we need to consider too many dynamical systems, the importance sampling adjustment fails to work in the high dimensional setting.  

In this paper, we propose an alternative way to share hierarchical data information. For each area, we generate a few auxiliary data points, combine them with the local data, and fit a local dynamical system to the combined data. The auxiliary data are chosen such that the posterior inference based on the combined local data is approximately the same as fitting the full hierarchical model across all areas. The auxiliary data are derived from relatively simple hierarchical models and have the same format as the originally observed data. As a result, the proposed model allows one sub-epidemic to effectively borrow the data information from other sub-epidemics without increasing the computational burden or the implementation cost of revising the dynamical models.

The rest of the paper is organized as follows. We introduce our motivating data in Section~2 and then the dynamic models used to produce national estimates in Section~3. In Section~4, we introduce our method that extends the sub-national estimation to include the hierarchical structure for sharing information across areas. In Section~5, we evaluate the model performance via cross-validation and present results for Nigeria and Thailand. In Section~6, we offer conclusions and discussion for future work. 



\section{HIV Surveillance Data}

HIV sentinel surveillance data refer to the systematic and periodic collection of HIV-related information from specific populations or locations to monitor trends in HIV prevalence (the proportion of people living with HIV in a population) over time. Surveillance sites are chosen based on factors such as the burden of HIV, the presence of key populations, geographic distribution, and accessibility. The distant surveillance sites also shed light on the geographic distribution of the epidemic. At each sentinel site, data are collected annually or every few years from a sample of the target population. The HIV testing results are summarized by the numbers of tested individuals and confirmed HIV-positive cases. Those longitudinal measures of the HIV-positive proportions from the same site allow the timely detection of the HIV prevalence changes over time. 

HIV epidemics are typically classified into two main categories: generalized epidemics, where HIV prevalence is consistently higher than 1\% in the general population, and concentrated epidemics, where HIV prevalence is consistently high among one or more specific high-risk groups, such as people who inject drugs, men who have sex with men, sex workers, and their clients, but remains relatively low in the general population. We used surveillance data from Nigeria and Thailand, which were reported between the early 1990s and early 2010s, as illustrative examples of generalized and concentrated epidemics, respectively. 

{\em Nigeria} consists of 36 states and the Federal Capital Territory Abuja, and has one of the largest HIV epidemics in the world. Some states, particularly in the southern part of the country, have higher HIV prevalence rates than the national average (1.4\%). Ten states were identified with HIV prevalence above 2\%, nine of which had a significant unmet need for HIV treatment and were at risk of being left behind if no action was taken \citep{Nigeria2021}. We use it to represent countries with high (\textgreater 1\%) HIV prevalence. In such countries, the HIV sentinel surveillance data are mainly based on HIV prevalence among pregnant women, as measured in antenatal clinics (ANC). Nigeria's ANC data started in 1992, and the data quality varies across states: the number of years for which ANC data is available ranges from 6 to 9; the number of clinic sites per state ranges from 2 to 8. The number of tested individuals in each area ranges from 3,482 to 10,789, with a median of 4,850.

Nearly all countries established ANC HIV surveillance in the early 1990s, making it the earliest and most consistently available source of information for the time trend of HIV prevalence. However, these data exhibit known biases – they include only sexually active women of fertile ages and are collected from a nonrandom selection of sentinel sites. To compensate for those biases and generate nationally representative estimates of HIV prevalence for all adults, national population-based surveys (NPBS) later included HIV testing results. ANC surveillance and NPBS together provide a full picture of the HIV prevalence in the general population of countries with a high burden of HIV epidemics. ANC determines the relative change of the HIV prevalence over time; NPBS calibrates the ANC-based prevalence trend by adding a constant so that the calibrated trend matches NPBS prevalence at the survey year.

For illustrating the method, we used state-level HIV surveillance data for Nigeria collated for the national UNAIDS HIV estimates process in 2013, previously published by \cite{Mahy2014}.


{\em Thailand} represents epidemics where the HIV prevalence is lower and largely concentrated among specific sub-populations. The Thai epidemic grew rapidly in the late 1980s when prevalence among people who inject drugs, female sex workers and their clients rose rapidly nationwide, shortly followed by growing prevalence among pregnant women. Thailand has an extensive HIV surveillance data set for key populations dating from 1989, when surveillance was first instituted in 14 provincial capitals. By 1991, all provinces collected prevalence data in multiple populations \citep{Weniger1991,Brown1994}. 
The current epidemic estimation for Thailand is stratified into multiple sub-populations and four geographic areas: Central (26 surveillance sites), North (17 surveillance sites), Northeast (19 surveillance sites), and South (14 surveillance sites). Here, we focus on the sub-population of indirect sex workers. Prevalence data are observed from 1989 to 2011. Among 23 data years and 76 surveillance sites, there were 36.0\% missing data, and the average number of tested individuals was 56,796 per area.

\section{The HIV estimation model}
\label{sect-EPP}
\begin{figure}[!ht]
\includegraphics[width=12cm]{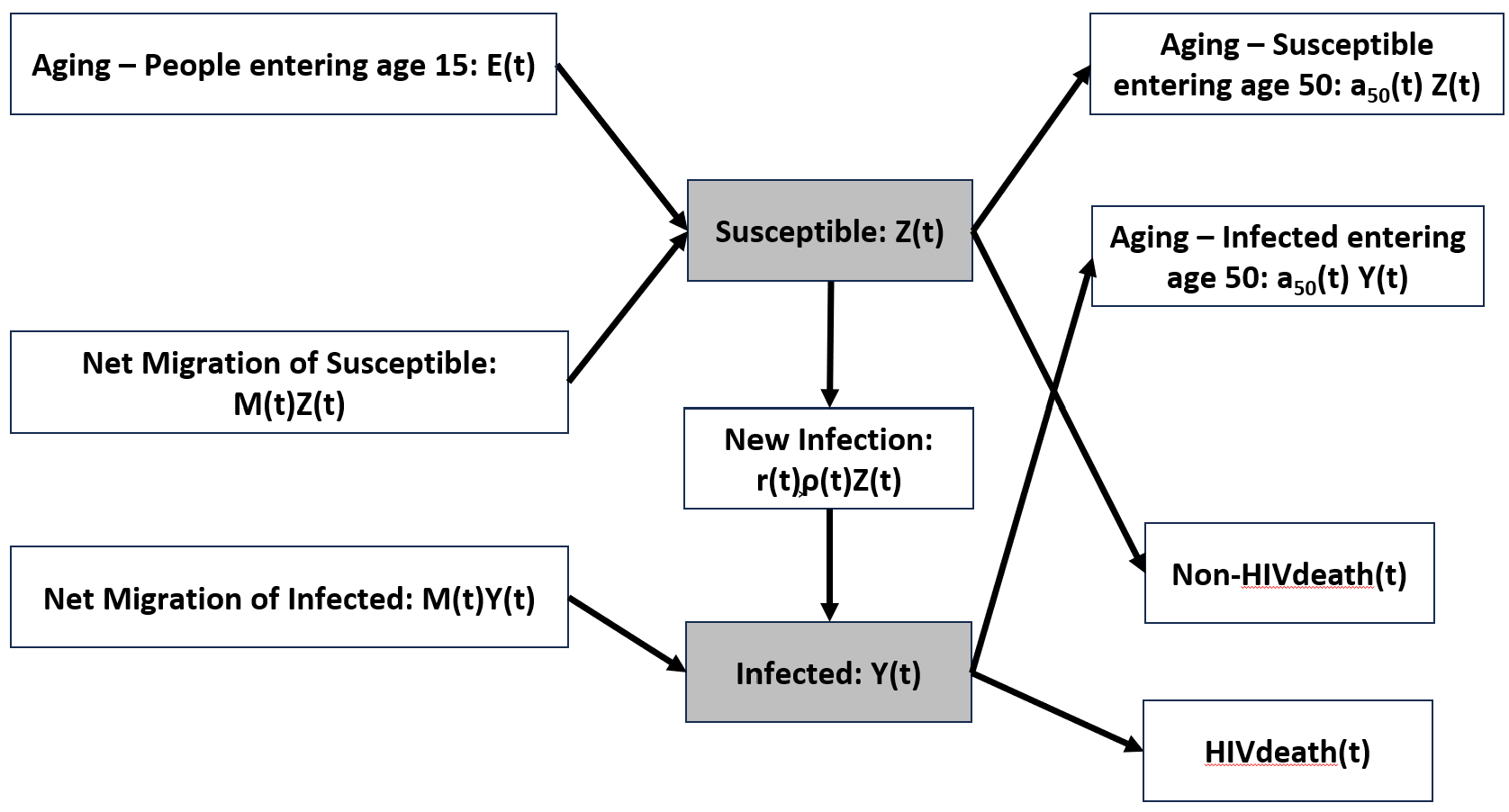}
\caption{\footnotesize{The EPP model compartments: The blue boxes represent the susceptible group (with size $S(t)$) and the infected group (with size $Y(t)$), both aged between 15 and 50. Susceptible individuals can be infected and move to the infected group. The shaded boxes represent the two compartments in the dynamic model, and the other boxes indicate the transitions.}}
\label{fig:EPP}
\end{figure}

The most commonly used model to estimate and project HIV epidemics is the Estimation and Projection Package (EPP), which is the primary source of incidence estimates in Spectrum; 170 countries have used it to estimate the impact of HIV on their populations in 2022. The EPP model is based on a susceptible-infected (SI) epidemiological model. The settings for the {\em general population} and the {\em high-risk groups} differ slightly. The basic dynamic model is as follows:
\begin{equation}
\left\{\begin{array}{ccc}
\frac{dZ(t)}{dt} & = & E(t) - r(t) \rho(t) Z(t) - \mu(t) Z(t) - a_{50}(t)Z(t) +  M(t)Z(t), \\
\frac{dY(t)}{dt} & = & r(t) \rho(t) Z(t) - \textup{HIVdeath}(t) - a_{50}(t)Y(t) +  M(t)Y(t).
\end{array}\right.
\label{eqn:ode}
\end{equation}
The model represents the adult (aged 15-49 years) population stratified into the susceptible population $Z(t)$ and infected population $Y(t)$ at time $t$, such that the total adult population size is $N(t) = Z(t) + Y(t)$. $E(t)$ is the number of new adults entering the population at age 15, $\rho(t)=Y(t) / N(t)$ is the prevalence, $\mu(t)$ is the non-HIV death rate, and $\textup{HIVdeath}(t)$ is the number of deaths among infected individuals. $a_{50}(t)$ is the rate at which adults exit the model after attaining age 50, and $M(t)$ is the rate of net migration into the population. External demographic data on the population define the quantities mentioned above, and most countries use annual estimates of fertility, non-AIDS mortality, and migration produced by the United Nations Population Division \citep{stover2017updates}. Figure \ref{fig:EPP} presents a diagram explaining the transitions between compartments of the EPP model. The UNAIDS reference group keeps updating the mortality rates of those living with HIV based on the current cohort studies \citep{stover2019updates}. Note that {\em high-risk group} members usually stay in that group for a certain period; thus, the high-risk group is not a closed population. The rates at which people enter and leave the high-risk group are estimated by biobehavioral surveys targeted at high-risk group members. Those rates are also fixed in the model.

The parameters that are needed to be estimated are involved in modeling the transmission rate from infected to uninfected adults. EPP posits that the logarithmic change in the infection rate is driven by three factors, expressed in the following manner (EPP r-trend model, \cite{Bao2012rtrend}):
\begin{eqnarray}
\log r(t+1)-\log r(t) &=& \beta_1 \times (\beta_0 - r(t)) + \beta_2 \rho(t) + \beta_3 \gamma(t), \nonumber \\
\gamma(t) &=& \frac{(\rho(t+1)-\rho(t)) (t-t_1)^+}{\rho(t)}, \label{eqn:rtrend}\\
(t-t_1)^+ &=&
\begin{cases}
t-t_1 & \text{for } t>t_1,\\
0 & \text{for } t\leq t_1. \nonumber 
\end{cases}
\nonumber
\end{eqnarray}
The parameter $\beta_0$ represents a stationary value of the infection rate at which the new infection rate offsets the mortality rate. The parameter $\beta_1$ characterizes the convergence rate of the infection rate towards $\beta_0$. Lastly, $\beta_2$ quantifies the relationship between the change in $r(t)$ and the prevalence. As discovered in \cite{Bao2012rtrend}, $\beta_1$ is generally estimated to be positive, while $\beta_2$ is typically estimated to be negative. The term $\gamma(t)$ represents the tendency for prevalence stabilization after year $t_1$. A negative value for $\beta_3$ suggests that the infection rate declines when the prevalence is too high and vice versa. Finally, $t_0$ denotes the epidemic's starting year when the prevalence is at 0.0025\%, and $r_0$ refers to the infection rate at $t_0$. Given a small initial seed prevalence at the start of the epidemic, the model simulates a temporal trend of prevalence, incidence, and HIV mortality rates.

As described in Section~2, the main source of data to estimate HIV prevalence trends has been sentinel surveillance data. These consist of the number of infected individuals $Y_{it}$ and the number of tested individuals $N_{it}$, for a given year $t$ and a given site $i$. We assume that the relative annual changes in HIV prevalence are the same between the ANC population and the general adult population (aged 15-49 years), with a time-invariant bias parameter, $\alpha$, characterizing any systematic difference between those two populations. To account for multiple sentinel sites in the same region, the observed clinic-level prevalence and the population prevalence -- $\rho(t)$ -- derived from the dynamical system (\ref{eqn:ode}) are linked through a mixed-effects model. The probabilistic model is summarized as follows:
\begin{eqnarray}
W_{it}&=&\Phi^{-1}(\rho_t)+\alpha+b_i+\epsilon_{it}, \nonumber\\
b_i&\sim& N(0,\sigma^2), \label{eqn:ranef} \\
\epsilon_{it}&\sim&N(0,\nu_{it}) \nonumber
\end{eqnarray}
where $\Phi^{-1}(.)$ is the inverse cumulative distribution function of the standard normal distribution, $W_{it}=\Phi^{-1}(\frac{Y_{it}+0.5}{N_{it}+1})$ is the sample proportion of HIV positive cases at the probit scale. The site-specific random effects, $b_i$'s, and the residuals, $\epsilon_{it}$ are all independent. The random effect variance, $\sigma^2$, is assumed to have an inverse-Gamma prior, which gets integrated out in the likelihood evaluation. The residual variance, $\nu_{it}$, is a fixed quantity that depends on the clinic data and approximates the binomial variation. For {\em generalized epidemics}, $(Y_{it}, N_{it})$'s may not be representative samples for the target population, i.e., ANC patients are biased samples for all adults (aged 15-49 years). We assume that the time trends of HIV prevalence at the probit scale between the sampling and target populations differ by a bias parameter, $\alpha$, which is time-invariant. 


A Bayesian framework is used to bring in the prior information of the unknown parameters, such as the starting time of the epidemic. Model parameters are estimated with posterior approximation via Incremental Mixture Importance Sampling (IMIS) \citep{Raftery2010}. The final estimates reflect three sources of information: the prevalence data, the prior knowledge, and the epidemic trends inferred by the SI model in Equation \ref{eqn:ode}. As a result, the EPP model has several advantages for estimating the HIV epidemics. Its basis in a dynamic epidemic model lends theoretical credibility to the resulting epidemic inference, and the model intrinsically represents the relationship between key epidemiological processes, enabling internally consistent estimates about multiple quantities of interest, including prevalence, mortality, and incidence, about which there is no directly observable data. Finally, the dynamic model is linked to a statistical model to produce all quantities of interest. 

\section{Incorporating hierarchical structure into estimating sub-epidemics}
\label{sect-Hierarchical models}
The current approach to extend the EPP model to fit the sub-national areas is to run the EPP model for each area independently using only data from within the area. However, many areas lack sufficient data for deriving stable and reliable results. 
Our goal is to improve the accuracy of the results in areas with sparse data while retaining the simplicity of the epidemic model and not increasing the computational burden. The main idea proposed here is very simple: we propose to fit a generalized linear mixed model (GLMM) with smoothing splines to data from all sub-national areas and incorporate the fitted prevalence into the EPP model as auxiliary data to approximate the hierarchical information. The approach can be outlined as follows:
\begin{enumerate}
\item For a particular country, we pool the data from all areas together and fit a GLMM with a non-parametric flexible time trend and random effects for areas and sites to represent the hierarchical structure.
\item We then use the predictive distribution of the area prevalence from the GLMM to create auxiliary data. 
\item Finally, we add the auxiliary data as pseudo sites to the original data and fit the EPP model separately for each area as before. The resulting prevalence, incidence, and mortality estimates can all be derived the same way as before. At the same time, they now contain information from other areas that would be included through the hierarchical structure.
\end{enumerate}
The details of the model are described in the following subsections.  

\subsection{Generalized linear mixed models (GLMM)}
In the first step, we model the prevalence data from all areas with  a GLMM so that the information can be pooled across areas in a computationally efficient way. 
Instead of using the dynamic model, we model prevalence trends as functions of time over the period in which data were observed. Spline models are highly flexible and customizable, enabling smooth and continuous curve fitting to represent nonlinear patterns and relieve over-fitting pressure \citep{Wold1974,Wahba1990}. We use the p-spline model with equally spaced knots to describe the overall time trend. We add area random effects to the spline coefficients to allow heterogeneity across areas. We fit the above with Stan implemented in the R package -- {\it brms} \citep{carpenter2017stan,burkner2017brms}. 

In a typical epidemic with multiple areas and multiple surveillance sites within each area, let $a$, $i$, $t$ indicate area, site, and time respectively. It is expected that prevalence data collected from the same site or different sites within the same area are correlated. We test GLMM specifications of varying complexity, and the following model is recommended:
\begin{equation}
\left\{\begin{array}{ccc}
Y_{iat} & \sim & \mbox{Binomial}(n_{iat},\rho_{iat}), \\
\mbox{logit}( \rho_{iat} )& = &\beta_{0} + b_{i(a)} +\sum_{d=1}^D \beta_d f_d(t) + \sum_{d=1}^D b_{ad} f_d(t),
\end{array}\right.
\label{eqn:GLMM1}
\end{equation}
where $f_{1}, f_{2}, \cdots, f_{D}$ are the spline basis functions; $\beta_{1}, \beta_{2}, \cdots, \beta_{D}$ are the spline coefficients with the penalty term, $\lambda \sum_{d=1}^{D-2} (\beta_{d+2}-2\beta_{d+1}+\beta_{d})^2$; $b_{i(a)}$ is the site-level zero-mean random intercept, and $b_{ad}$ is the area-level zero-mean random effect for the spline coefficient. 
We use weakly informative normal priors with a mean of zero and a variance of 100 for the fixed effects. We use student-t distribution with 3 degrees of freedom, a mean of zero, a scale of 2.5, and truncated for the positive part for the random effect variances.

We use the Hamiltonian Monte Carlo implemented in the R package -- {\it brms} \citep{burkner2017brms}, to approximate the posterior distribution of area-specific prevalence, $\rho_{at}$, with the site effects being marginalized. The parameters $\lambda$ and the variances of the random effects are also estimated in {\it brms} package. We run four separate sequences of Markov Chain Monte Carlo (MCMC) chains with 2,000 iterations for each chain to ensure convergence and robustness in the estimates. The convergence of Markov chains is checked based on Gelman-Rubin diagnostic ($\hat{R}>0.99$) and the effective sample size (ESS $>2000$).

\subsection{Incorporating hierarchical information into EPP model estimation}
GLMM utilizes data information efficiently by assuming similarity of the time trends (spline coefficients) while allowing heterogeneity across areas so that local data and data from other areas will jointly determine the prevalence trend in a specific area. However, it ignores the epidemic model and only provides inference for HIV prevalence trends within the period in which data were observed. One could have replaced the spline function with the prevalence trend produced by the epidemic model in Equation \ref{eqn:ode}. The difficulty is that estimating parameters in dynamic models is time-consuming as the posterior distribution tends to be concentrated around nonlinear ridges and can also be multimodal (\cite{Raftery2010}). Estimating multiple dynamical systems jointly will increase the computing cost and make it infeasible for a country with many areas.

Instead, we take the GLMM posterior estimates for the area prevalence and add them to the EPP model as auxiliary data for each area. Let $P_{\textup{full}}(\rho_{at})$ be the posterior distribution of the HIV prevalence for area $a$ and year $t$ based on the full hierarchical model. The vector $\rho_a = \{\rho_{at}\}$ has the posterior mean, $\mu_a$, and the posterior covariance $\Sigma_a$. We convert this information for area $a$ to a series of binomial observations with prevalence $\mu_{at}$ and sample size $n_{at}=K$ for $t$ in the data range. This binomial formulation is the same as the original data and thus facilitates the straightforward inclusion of auxiliary information into the existing EPP likelihood function. 

The amount of information the auxiliary data carries is controlled by the sample size $K$. As $K$ increases, the auxiliary data become more informative, and $K=0$ corresponds to fitting Spectrum/EPP to data within area $a$ without borrowing information from other areas. To choose the proper value of $K$ for the area $a$, we refit the penalized spline model to area $a$'s data augmented by the auxiliary data of sample size $K$ and refer to it as the augmented data model. Let $Q_K(\rho_{at})$ be the resulting posterior distribution of area-level HIV prevalence. It has posterior mean $\mu_{aK}$ and posterior covariance $\Sigma_{aK}$. Assuming that both $P(\rho_{at})$ and $Q_K(\rho_{at})$ approximately follow multivariate normal distributions, we calculate their Kullback-Leibler (KL) divergence as follows:
\[
2 D_{\textup{KL}}(P|Q_K) =  \textup{tr}(\Sigma_{aK}^{-1} \Sigma_a) - \textup{dim}(\Sigma_a) + (\mu_{aK} - \mu_a)^T 
\Sigma_{aK}^{-1}(\mu_{aK} - \mu_a) + \ln{\textup{Det}(\Sigma_{aK})} -
\ln{\textup{Det}(\Sigma_{a})}.
\]
For a set of distinct values of $K$, we choose the one that leads to the smallest KL divergence. 

\subsection{Model validation}
We evaluate the model performance via cross-validation as outlined below:
\vspace{-0.05in}
\begin{enumerate}
\item For each surveillance site, randomly assign observations to the training and the test sets with a ratio of 1:1. So, we always have training data at each site to estimate the site-specific random effect for prediction purposes. We do not further stratify the training test split by years as we want to use simple random sampling to mimic all potential missingness patterns that could arise.
\vspace{-0.08in}
\item Apply a full hierarchical model in Equation \ref{eqn:GLMM1} to the training data of all sub-epidemics.
\vspace{-0.08in}
\item Generate auxiliary binomial data from the predictive distributions of GLMM with different sample sizes: $K=25, 50, 100, 200, 400, 800, 1600, 3200$, and choose the optimal sample size that minimizes the KL divergence between the augmented data model and the full hierarchical model.
\vspace{-0.08in}
\item We compare the observations in the test data between the augmented data model and the independent model that does not use auxiliary data. We consider the following evaluation metrics: (1) the mean absolute error (MAE) is the absolute difference averaged over all surveillance sites and years in the test set, (2) the coverage is the proportion of observed data in the test set being covered by the corresponding 95\% predictive intervals, (3) and the average width of the 95\% predictive intervals.
\end{enumerate}

\section{Results}
\label{sect-Results}
We illustrate the model performance with Nigeria and Thailand examples, one representing generalized epidemics and the other representing concentrated epidemics. 
The R code for EPP implementation and the data augmentation is available at \url{https://github.com/lebao0215/Dynamic-Models-Augmented-by-Hierarchical-Data}. 
We add random noise to the datasets and the most updated datasets shall be requested through \url{https://hivtools.unaids.org/spectrum-file-request/}.

\subsection{Nigeria Example}
We first apply the cross-validation procedure to Nigeria's 37 sub-national areas. The training data contains half the full data to induce data sparsity, and surveillance sites stratify the training-test splits to estimate the site-specific effects. We repeat cross-validation under ten random training-test splits and summarize the average results in Table \ref{tab:1} (a). With the augmented data, we reduced the mean absolute error (MAE) from 2.39 to 2.13 as measured at the original scale of the HIV prevalence in percentage. We also substantially reduced the width of the 95\% prediction interval from an average of 7.89 for the independent model to an average of 6.45 for the augmented data model. It came with a cost of slightly decreased coverage of the prediction intervals, from 84.6\% to 82.0\%. The relative changes, defined as the new value minus the old value and then divided by the old value, are $-10.9\%$, $-18.3\%$, and $-3.1\%$, for MAE, Width and Coverage, respectively. We also calculate the average computing time with a 2.2 GHz Intel Xeon Processor and 128 GB RAM server. It took 16.3 minutes to apply the EPP model to one area's surveillance data and 24.0 minutes to apply the EPP model to both one area's surveillance data and the augmented data. In addition, running each GLMM took 1.5 minutes. So, selecting the optimal sample size and creating the augmented data add 22.5 minutes to the total computing cost.

\begin{table}[!h]
    \centering
    \caption{A cross-validation comparison between the independent model and the augmented data model. The mean absolute error (MAE) is measured in the percentage of HIV prevalence. Similar to MAE, the coverage and the width of 95\% credible intervals are averaged over all 370 test datasets, which include 4,430 test observations in total.} 
    \begin{tabular}{lcccc}
    \toprule
         \multicolumn{5}{c}{(a) Nigeria antenatal clinic patients}  \\
        \hline
         &  MAE  & Width  & Coverage & EPP Time \\
    \hline
        Independent Model & 2.39 & 7.89 & 84.6\% & 16.3 min (EPP)\\
    \hline
        Augmented Data Model & 2.13 & 6.45 & 82.0\% & 24.0 min (EPP) $+$ 22.5 min (GLMM) \\
    \hline 
    \midrule
         \multicolumn{5}{c}{(b) Thailand female sex workers}  \\
    \hline
         &  MAE  & Width  & Coverage & Computing Time \\
    \hline
        Independent Model & 4.44 & 11.76 & 81.2\% & 33.5 min (EPP)\\
    \hline
        Augmented Data Model & 2.53 & 7.43 & 80.7\% & 44.3 min (EPP) $+$ 11.6 min (GLMM)\\
    \bottomrule
    \end{tabular}
    \label{tab:1}
\end{table}

\begin{figure}[!ht]
\includegraphics[width=14cm]{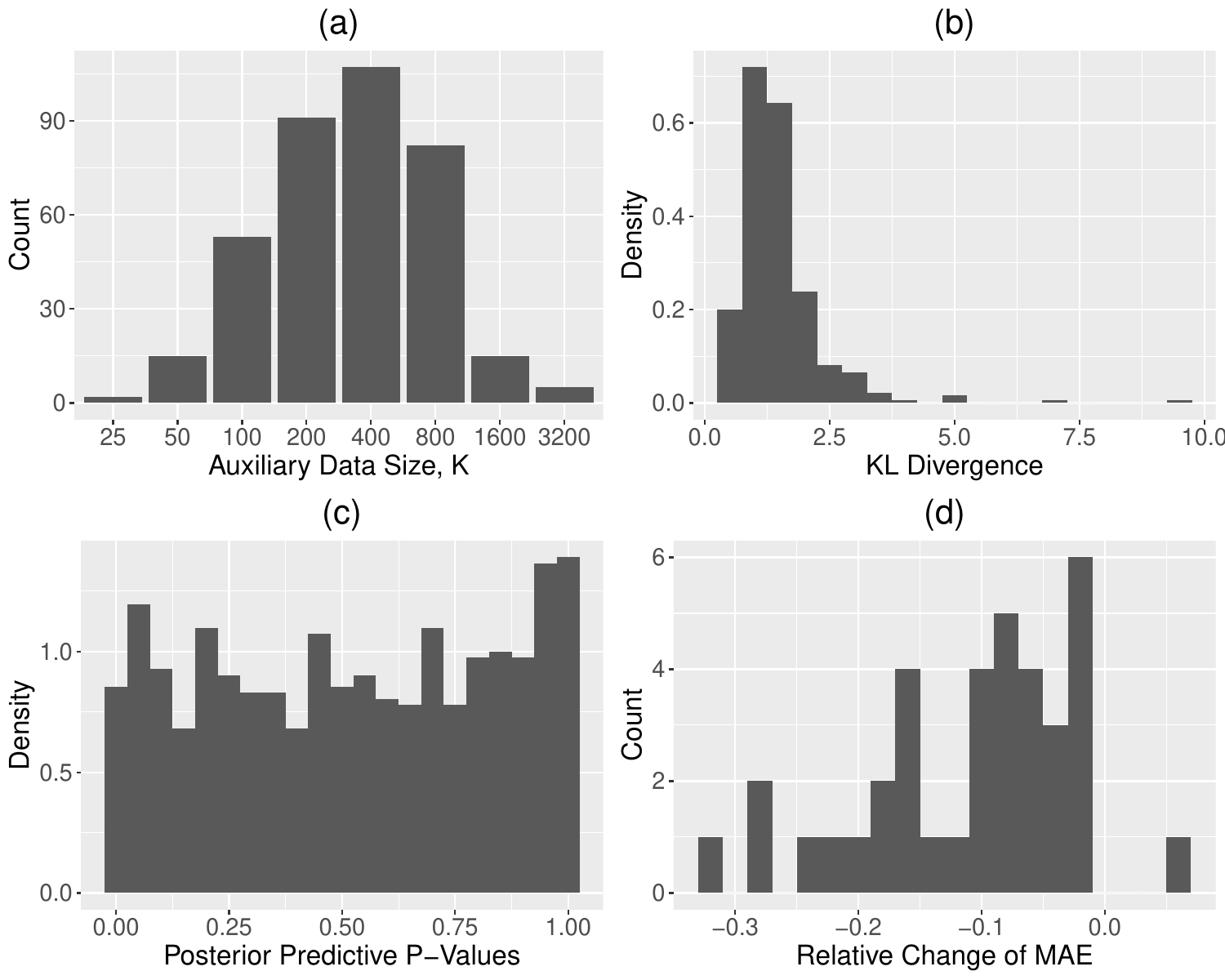}
\caption{\footnotesize{Nigeria Training-Test Splits: (a) the histogram for the selected auxiliary sample size based on training datasets; (b) the histogram of KL divergence between the hierarchical model posterior and the single area model augmented by the auxiliary data; (c) the histogram of the quantile of observed prevalence under the posterior predictive distribution; (d) the histogram of test data MAE by areas.}}
\label{fig:Nigeria1}
\end{figure}

Figure \ref{fig:Nigeria1} (a) presents the distribution of selected auxiliary data sample size, $K$, and Figure \ref{fig:Nigeria1} (b) shows the distribution of corresponding minimum KL divergence. Most areas took $K=200 \sim 800$, similar to the sample sizes of the real surveillance data, and it rarely picked $K=25$ or $3,200$. For those selected $K$, all KL divergences were less than 10, with the majority (99.2\%) less than 5. It indicates that using one set of auxiliary data can be as effective as using datasets from all other areas regarding information sharing.
We conduct the posterior predictive check by examining the quantiles of observed data points given their posterior predictive distributions. It evaluates the adequacy of a fitted model with respect to the observed data, and an empirical distribution of the posterior predictive p-values close to a uniform distribution is desirable. 
Figure \ref{fig:Nigeria1} (c) shows that our posterior predictive p-values approximately follow the uniform distribution between 0 and 1. There are a few more cases with large values because of discrete values of the $y_{iat}$, which is the count of HIV-positive individuals.
For each area, we calculate the relative change of MAE from the independent model to the augmented data model. The negative relative change indicates an improved prediction accuracy. Figure \ref{fig:Nigeria1} (d) provides the histogram of the relative changes. There are 16 areas with relative change less than $-10\%$ where the data augmented model substantially improved prediction accuracy, 20 areas with relative changes between $-10\%$ and $0$ and thus mild improvements, one area with a positive relative change, 6.4\%, indicating the augmented data model performed slightly worse than the independent model.

\begin{figure}[!ht]
\begin{tabular}{cc}
\begin{minipage}{7cm}
\includegraphics[width=7cm]{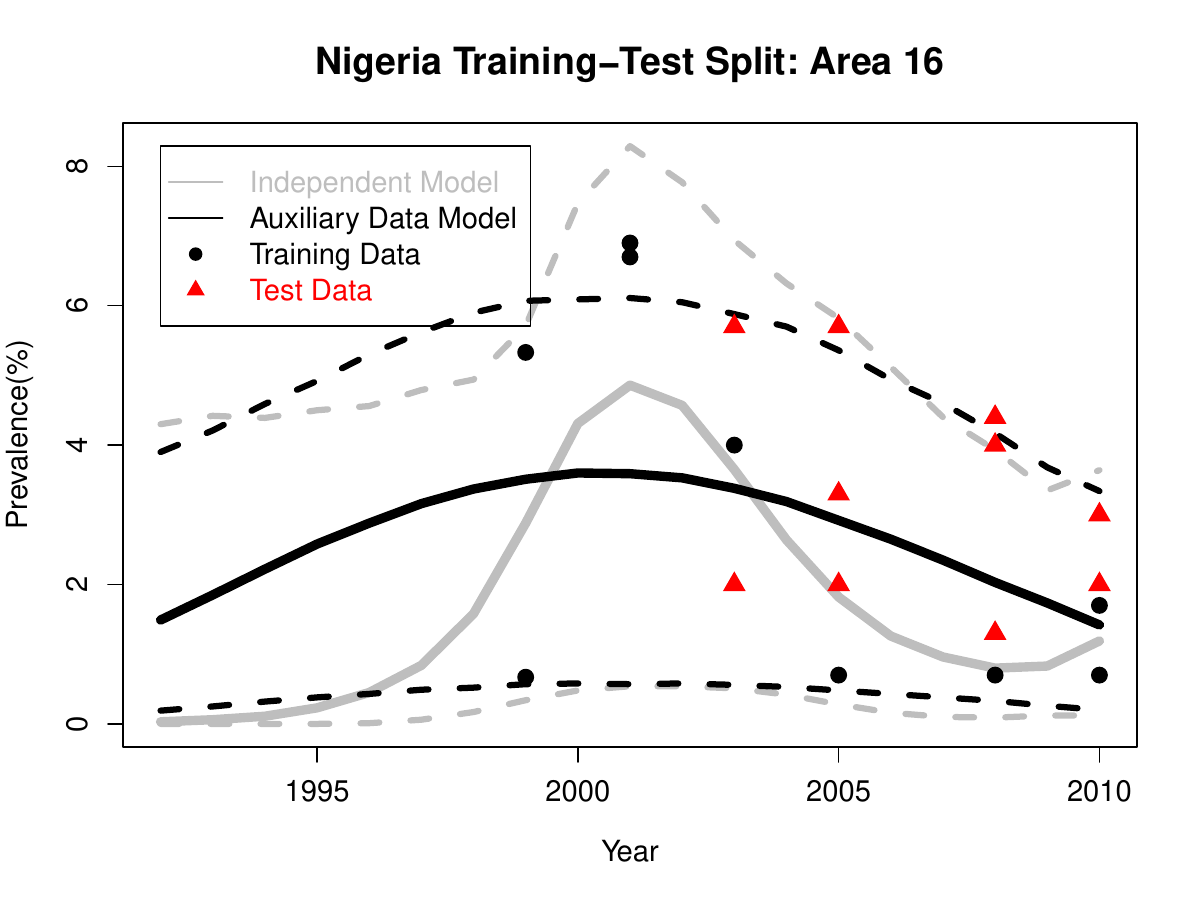}
\end{minipage}
&
\begin{minipage}{7cm}
\includegraphics[width=7cm]{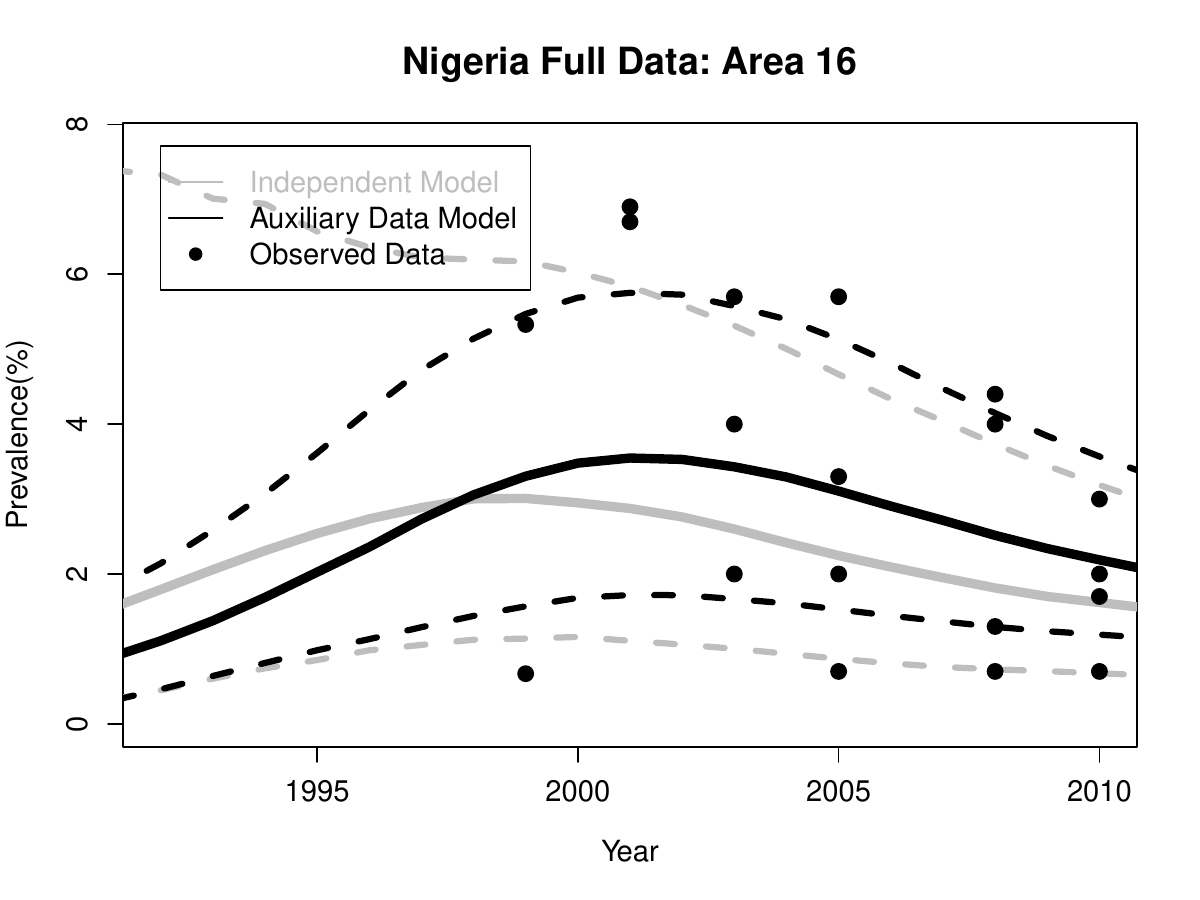}
\end{minipage}
\\
\end{tabular}
\caption{\footnotesize{Nigeria's HIV prevalence estimation. 
The left panel shows the results of fitting the EPP model to the training data, and the test data are shown in red; the right panel shows the results of fitting the EPP model to the full data.
The gray curves show the posterior median and 95\% credible interval of prevalence trends estimated from the original EPP model without using auxiliary data; the black curves show the posterior median and 95\% credible interval of prevalence trends estimated from EPP augmented by auxiliary data; the black dots show the observed training data or full data.}}
\label{fig:Nigeria16}
\end{figure}

Figure \ref{fig:Nigeria16} compares the estimated area-level prevalence trends from different models in one area as an illustrative example. The gray curves show the posterior median and 95\% credible interval of prevalence trends estimated from EPP without using auxiliary data. In the training-test split scenario (left panel), the gray curves indicate a quick increase in HIV prevalence around 2000, followed by a quick decline in the late 2000s, consistent with the training data (black dots). 
The black curves show the posterior median and 95\% credible interval of prevalence trends estimated from EPP with auxiliary data.
It leads to a more steady change in HIV prevalence, which agrees better with the test data (red dots) and the full data result (right panel). For the other areas, we present the full data analysis results in Appendix A.

\subsection{Thailand Example}
Thailand is among the countries with the highest quality surveillance data for key sub-populations that have high-risk behaviors. In a typical country with a concentrated HIV/AIDS epidemic, surveillance data for key sub-populations are often very sparse due to the stigmatized nature of those sub-populations, and there may not be sufficient data from a single high-risk group for a reliable estimation. We want to mimic the sparse data situation when constructing the training set. For each sub-population in each area, we randomly select three sites and then take three years of data from each site as the training data and use the rest as the test data. Following the procedure described in Section 4.3, we repeat the random training-test splitting and corresponding evaluations ten times.

\begin{figure}[!h]
\includegraphics[width=16cm]{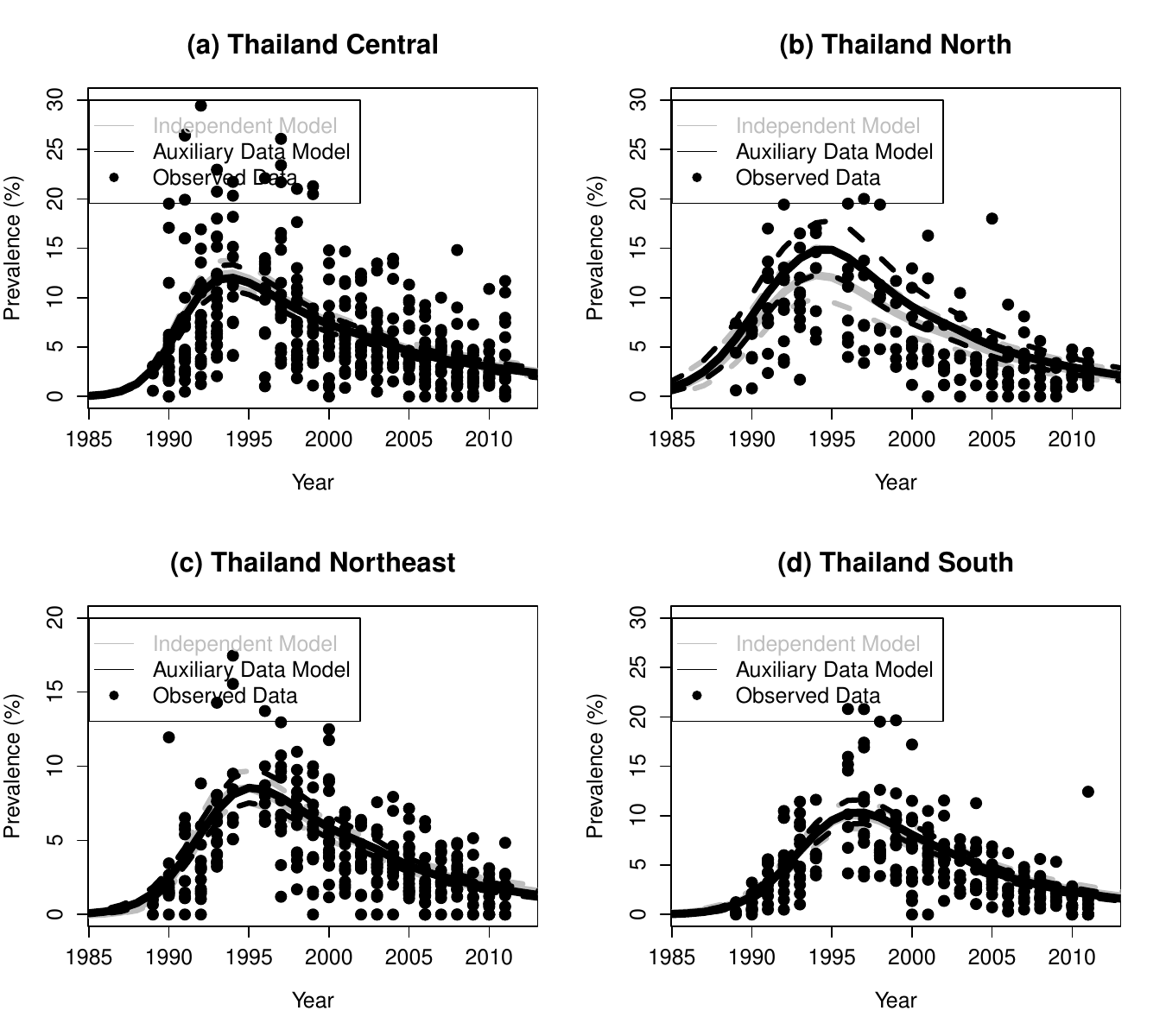}
\caption{\footnotesize{Thailand indirect sex worker example. The gray curves show the posterior median and 95\% credible interval of prevalence trends estimated from the original EPP model without using auxiliary data; the black curves show the posterior median and 95\% credible interval of prevalence trends estimated from EPP augmented by auxiliary data; the black dots show the observed data.}}
\label{fig:Prev_ThaiFull}
\end{figure}

Table \ref{tab:1} (b) shows that with the augmented data, we reduced the mean absolute error (MAE) from 4.44 to 2.53, a 43\% relative reduction, as measured at the original scale of the HIV prevalence in percentage; reduced the width of the 95\% prediction interval from 11.76 to 7.43, a 37\% relative reduction; and slightly decreased coverage of the prediction intervals, from 81.2\% to 80.7\%. Running the GLMM, selecting the auxiliary data sample size, and fitting the EPP model with augmented data increased the total computing time from 33.5 minutes (running the original EPP model) to 55.9 minutes.
Finally, Figure \ref{fig:Prev_ThaiFull} shows the estimated prevalence trends among indirect sex workers in four regions of Thailand when the full datasets are used. The estimated prevalence trends are similar between EPP models with (black) and without (gray) using auxiliary data. 


\section{Discussion}
\label{sect-Discussion}
In this article, we describe an innovative approach that allows the sharing of data information across complicated dynamical systems. The proposed method strengthens the estimation of HIV epidemics at sub-national levels, which is critical for local program planning, decision-making, and resource allocation. We demonstrated that our simple, pragmatic approach of generating auxiliary data from nearby regions could improve both the accuracy and precision of sub-national HIV estimates without increasing the computational burden. This method will enable reliable and routine model-based estimates at more granular levels, information urgently needed for guiding and evaluating HIV policy.

The general approach proposed here lends itself to further extensions. We have chosen the penalized splines to model the flexible time trend. Other non-parametric models or time series methods are also possible. The spatial dependence could be considered by specifying a conditional auto-regressive (CAR) error structure for the area-level effects. In addition, one could consider including predictors in the GLMM, such as population density, average income, and proportion of migrants. In this article, we use a set of relatively simple models so that the audience can focus on the key ideas: separating the data-sharing part from the dynamical systems and incorporating the hierarchical structure through the auxiliary data. 

We note that there is often a high proportion of missingness in the HIV surveillance data. First, the frequency of HIV surveillance varies by country and depends on available resources and public health priorities. Some countries perform routine HIV surveillance annually, while others may do so every two to five years. We consider the missingness during surveillance gap years as missing at random (MAR). Second, the surveillance sites have different starting years. The new sites are chosen based on factors such as prioritizing the areas with a high burden of HIV, ensuring geographic coverage, and collaborating with existing health facilities. We assume the site selection bias has been a constant over the years. Thus, the site-specific random effect in the mixture model mitigates the impact of site selections on trend estimation. Third, some antenatal clinics were closed for renovation, and patients who would have gone to this clinic would likely go to neighboring clinics when it was closed. It will be interesting to further investigate the impact of such missingness, but it is beyond the scope of the current study as it requires additional knowledge of the reasons for the missingness.

The method described in this paper has been implemented into EPP/Spectrum as an important new feature \citep{niu2017incorporation}. The UNAIDS Reference Group on Estimates, Modelling and Projections continues to improve and refine the EPP models. Some recent advances include incorporating routine testing data, which gradually replaced the unlinked anonymous testing data in high HIV prevalence countries \citep{sheng2017statistical}, accounting for non-sampling errors \citep{eaton2017accounting}, introducing age-sex specific model \citep{eaton2019estimation}, and updating disease progression and mortality with untreated HIV infection \citep{glaubius2021disease}. Some other directions of statistical modeling include stochastic versions of the compartment model, taking the migration into account, and incorporating more information from the surveys and program data. Our proposed model is compatible with all variants of EPP models as it compresses the transferred information from other sub-epidemics into pseudo data points, and such a transfer learning process does not depend on the specifications of the dynamic model. The method is generic enough to model the time trends of other indicators, such as incidence and mortality, upon data availability. It can also be extended to various applications, such as cell movements, weather forecasting and demography.

\section*{Disclaimer}
The results presented in this paper are based on illustrative HIV prevalence data, which may not be complete. As the development of the EPP package continues, the most recent version includes additional options for the underlying epidemic models and new features such as stratification by age and sex \citep{eaton2019estimation} and shifting the definition of sub-national areas. Our implementation does not incorporate those new features but instead is based on a relatively simple version of EPP with R code available at \url{https://github.com/lebao0215/Dynamic-Models-Augmented-by-Hierarchical-Data}. 
Therefore, our results should not be seen as replacing or competing with official estimates regularly published by countries and UNAIDS.

\section*{Acknowledgments}
This research was supported by the Joint United Nations Programme on HIV/AIDS and NIH/NIAID R01AI136664. The authors thank Mary Mahy, Keith Sabin, Wiwat Peerapatanapokin, and Timothy Hallett for their helpful discussions and data sharing.

\bibliographystyle{biorefs}
\bibliography{EPP}

\end{document}


\title{Appendix for ``Dynamic Models Augmented by Hierarchical Data: An Application Of Estimating HIV Epidemics At Sub-National And Sub-Population Level"}

\author{LE BAO$^\ast$ and Xiaoyue Niu\\[4pt]
\textit{Department of Statistics, Penn State University, University Park, PA, USA}\\[2pt]
Tim Brown\\[4pt]
\textit{Population and Health Studies, East-West Center, Honolulu, HI, USA}\\[2pt]
Jeffrey W. Eaton\\[4pt]
\textit{MRC Centre for Global Infectious Disease Analysis, School of Public Health, Imperial College London, London, UK}\\[2pt]
{lebao@psu.edu}}

\markboth%
{L.Bao et. al.}
{Dynamic Models Augmented by Hierarchical Data}

\maketitle

\footnotetext{To whom correspondence should be addressed.}
\maketitle

\appendix
\makeatletter   
 \renewcommand{\@seccntformat}[1]{APPENDIX~{\csname the#1\endcsname}.\hspace*{1em}}
 \makeatother
 
\section{Full Data Analysis Results For Nigeria}
In Nigeria, we model the HIV prevalence among pregnant women in 37 areas, present the full data analysis result for one area in the main manuscript and present the results for the rest of the areas here. Some fitted curves do not go through the observed antenatal clinic data because they are pulled away by the HIV prevalence estimated in the national survey.

\begin{figure}[!h]
\includegraphics[width=16cm]{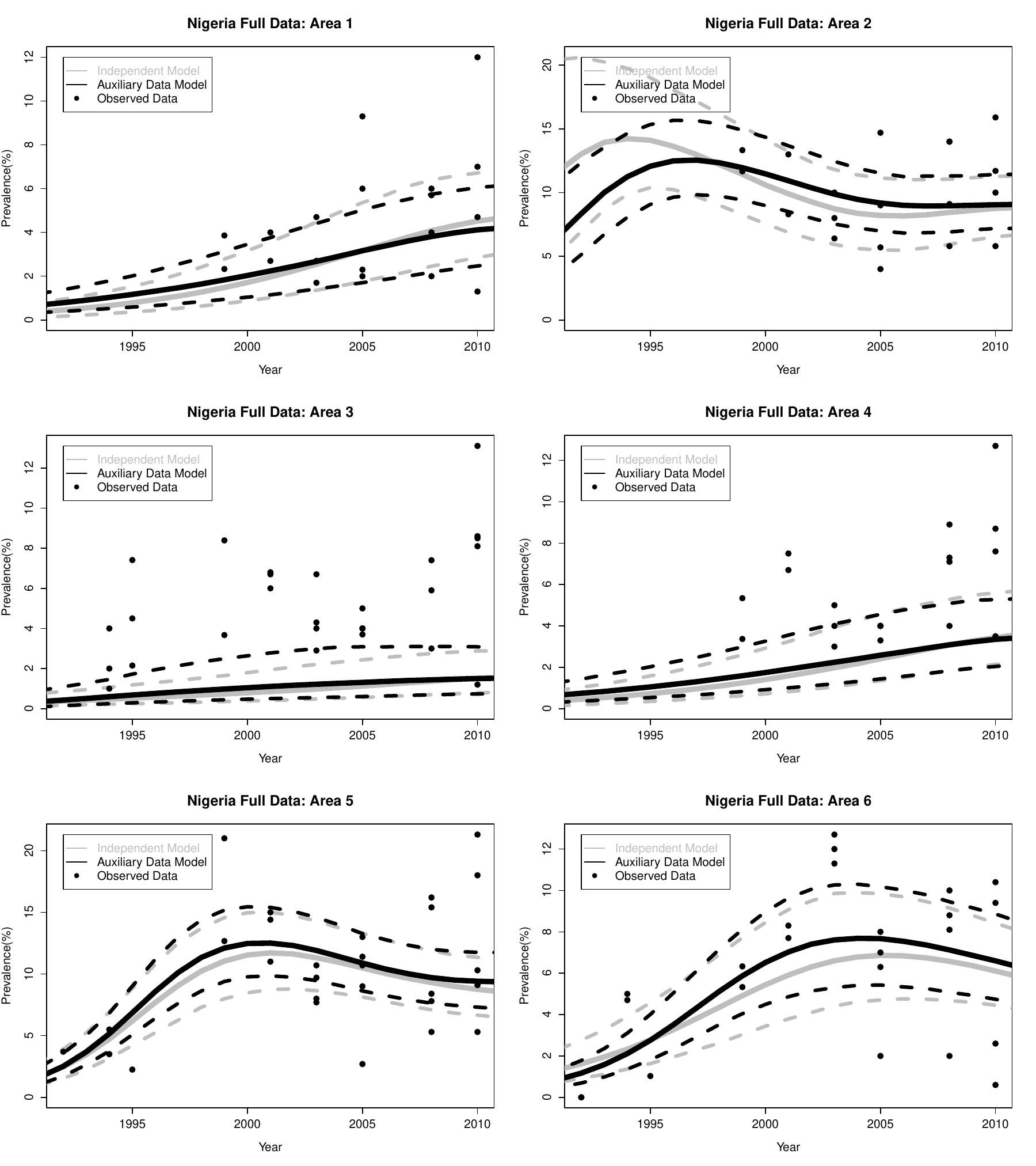}
\caption{\footnotesize{Full Data Analysis Results For Nigeria. The gray curves show the posterior median and 95\% credible interval of prevalence trends estimated from the original EPP model without using auxiliary data; the black curves show the posterior median and 95\% credible interval of prevalence trends estimated from EPP augmented by auxiliary data; the black dots show the observed data.}}
\end{figure}

\begin{figure}[!h]
\includegraphics[width=16cm]{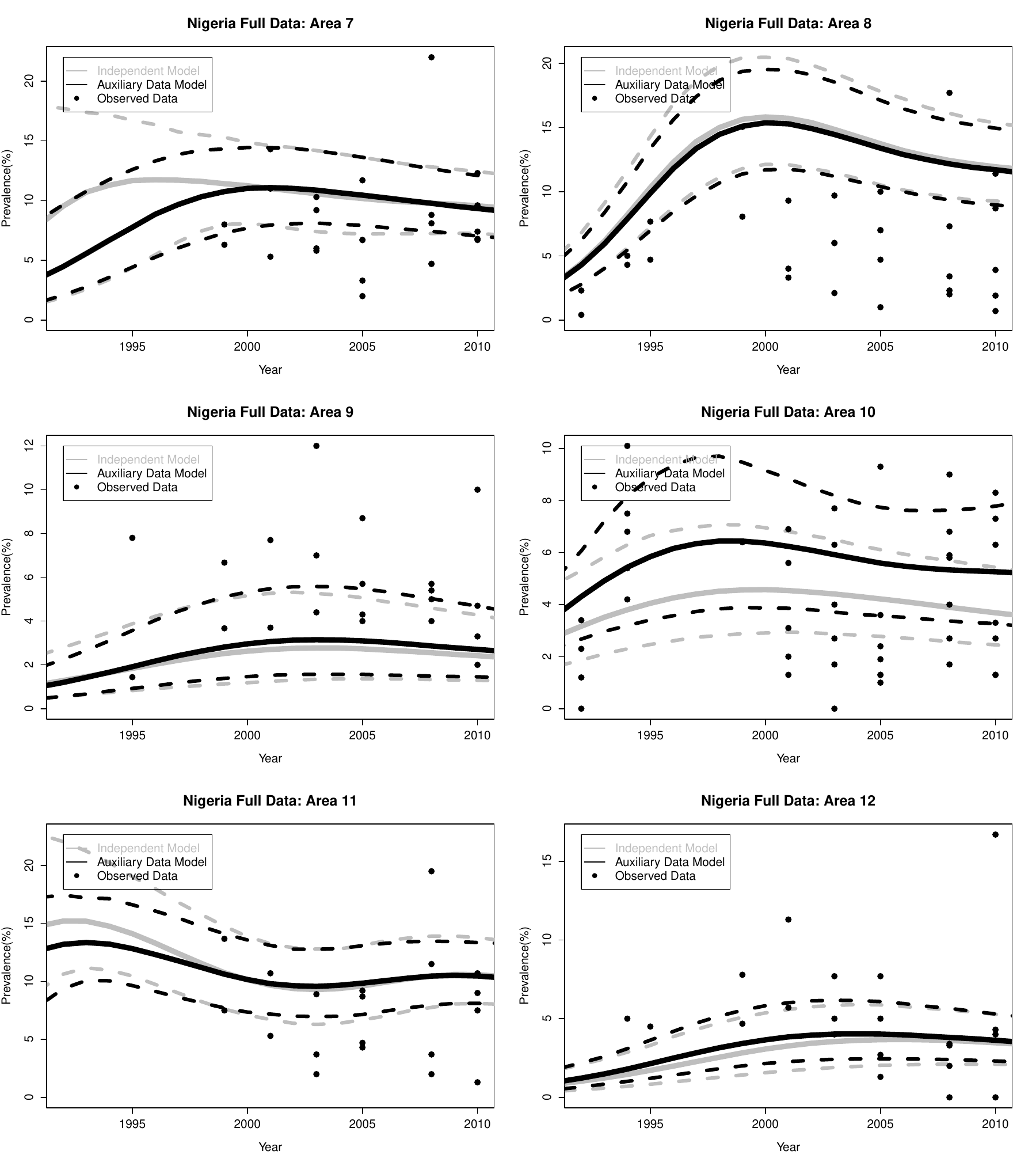}
\caption{\footnotesize{Full Data Analysis Results For Nigeria. The gray curves show the posterior median and 95\% credible interval of prevalence trends estimated from the original EPP model without using auxiliary data; the black curves show the posterior median and 95\% credible interval of prevalence trends estimated from EPP augmented by auxiliary data; the black dots show the observed data.}}
\end{figure}

\begin{figure}[!h]
\includegraphics[width=16cm]{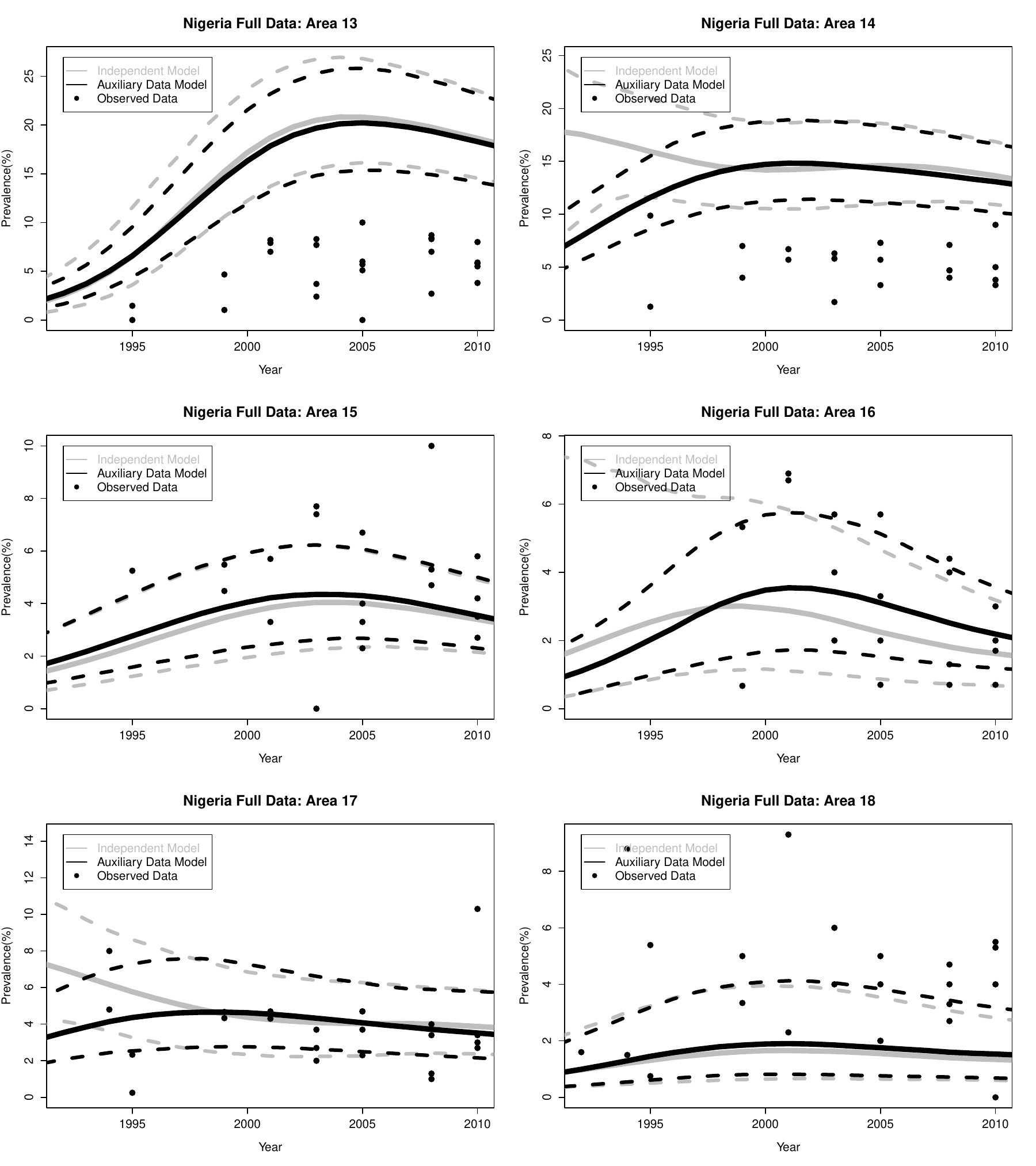}
\caption{\footnotesize{Full Data Analysis Results For Nigeria. The gray curves show the posterior median and 95\% credible interval of prevalence trends estimated from the original EPP model without using auxiliary data; the black curves show the posterior median and 95\% credible interval of prevalence trends estimated from EPP augmented by auxiliary data; the black dots show the observed data.}}
\end{figure}

\begin{figure}[!h]
\includegraphics[width=16cm]{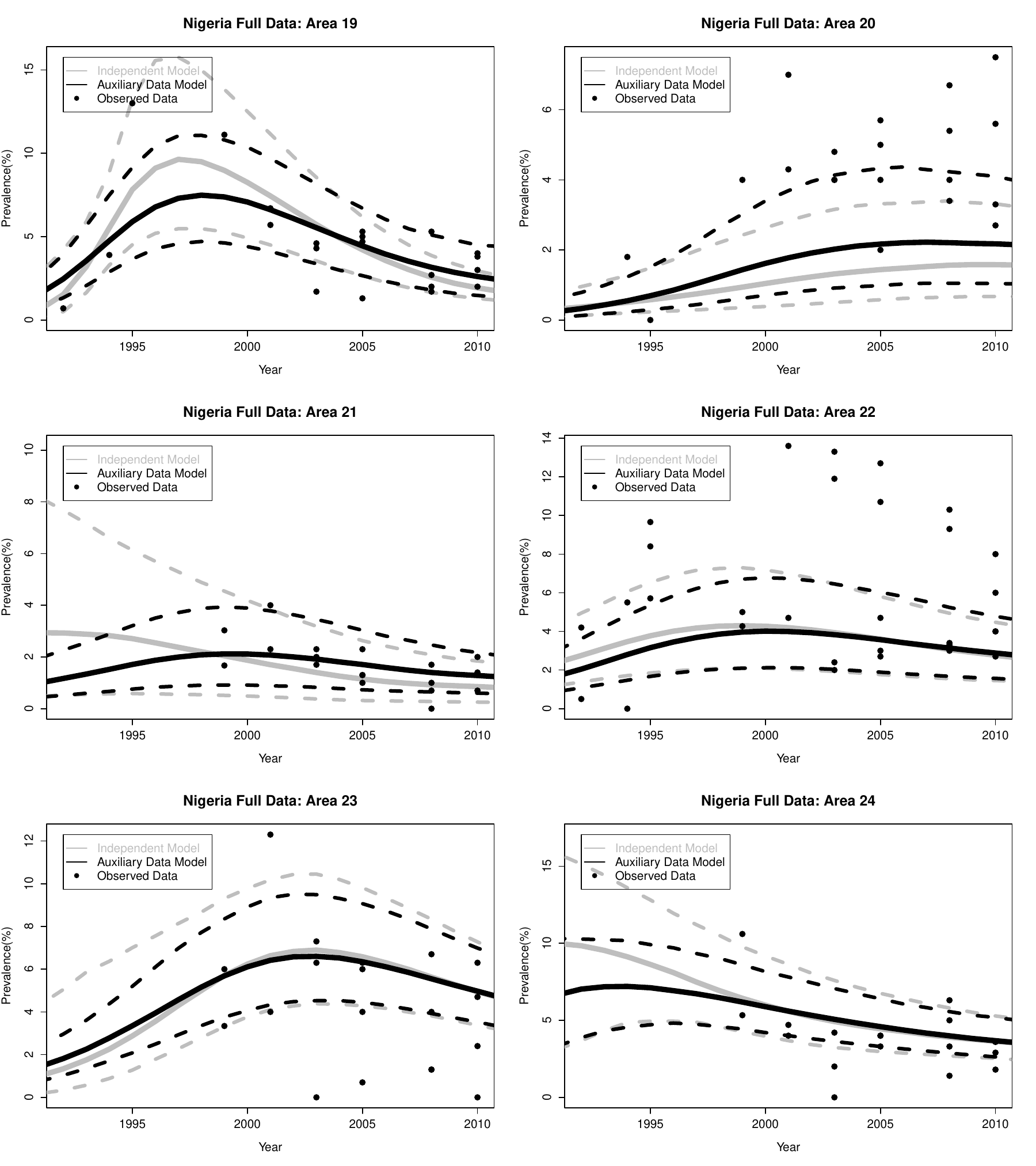}
\caption{\footnotesize{Full Data Analysis Results For Nigeria. The gray curves show the posterior median and 95\% credible interval of prevalence trends estimated from the original EPP model without using auxiliary data; the black curves show the posterior median and 95\% credible interval of prevalence trends estimated from EPP augmented by auxiliary data; the black dots show the observed data.}}
\end{figure}

\begin{figure}[!h]
\includegraphics[width=16cm]{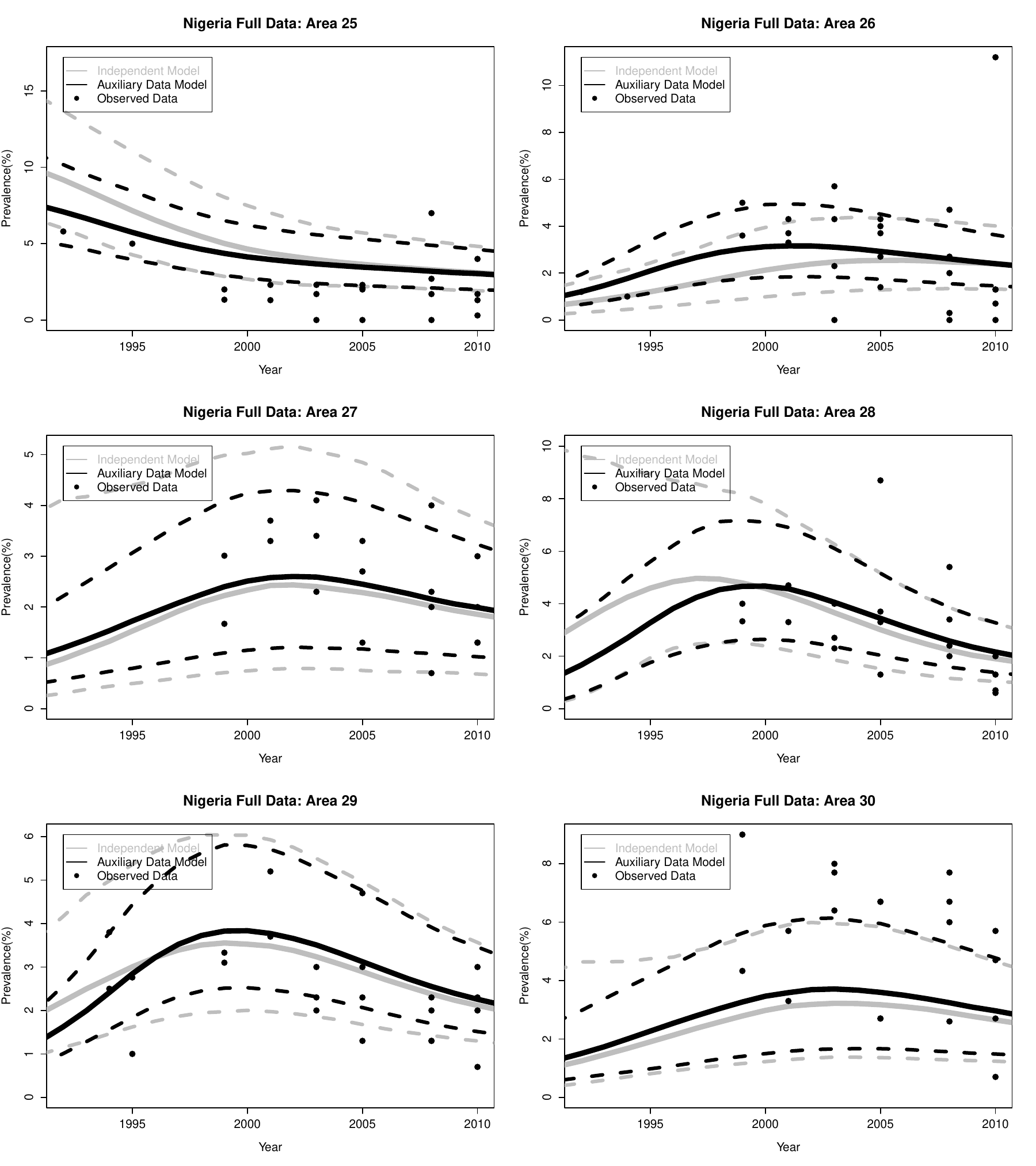}
\caption{\footnotesize{Full Data Analysis Results For Nigeria. The gray curves show the posterior median and 95\% credible interval of prevalence trends estimated from the original EPP model without using auxiliary data; the black curves show the posterior median and 95\% credible interval of prevalence trends estimated from EPP augmented by auxiliary data; the black dots show the observed data.}}
\end{figure}

\begin{figure}[!h]
\includegraphics[width=16cm]{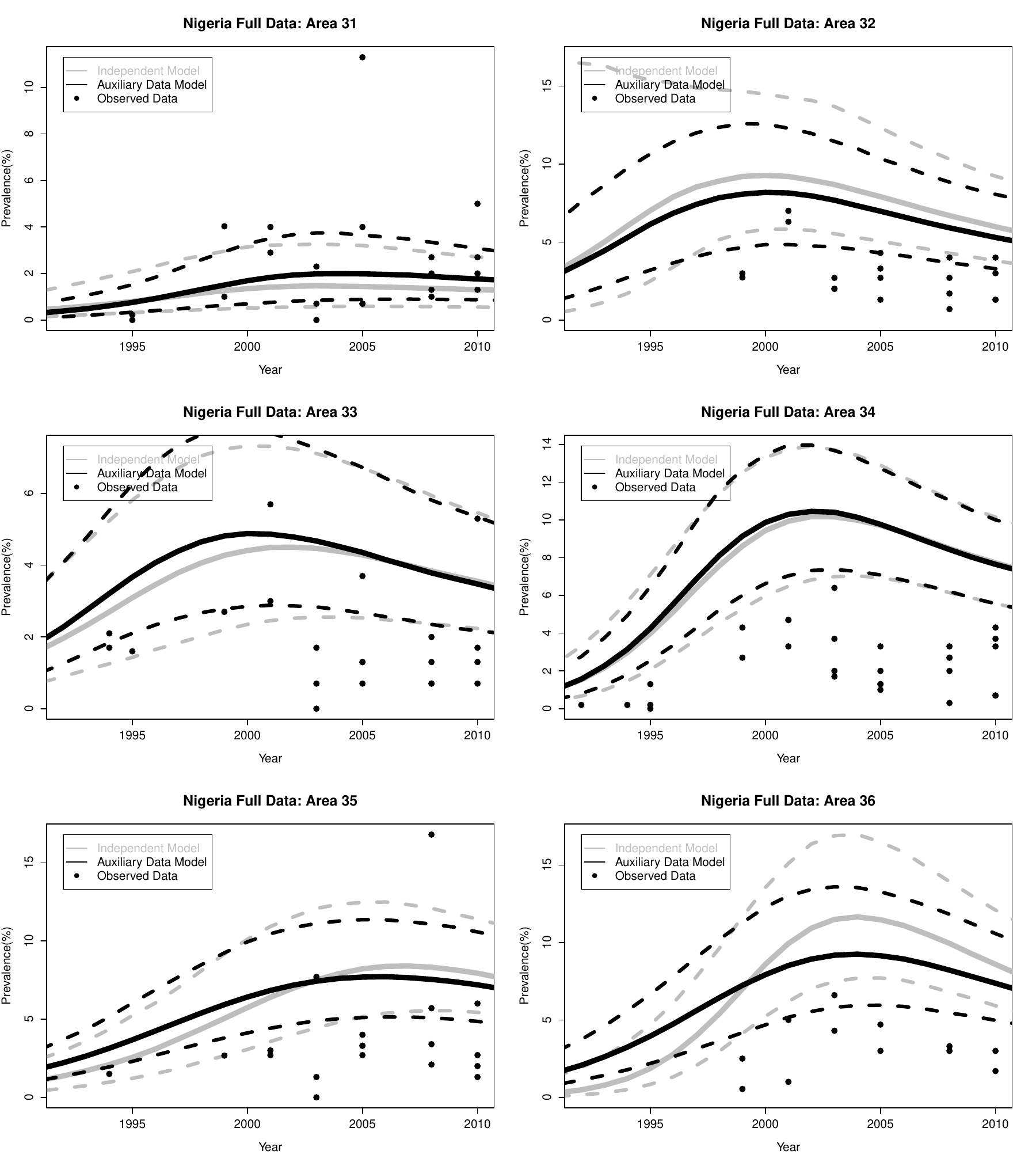}
\caption{\footnotesize{Full Data Analysis Results For Nigeria. The gray curves show the posterior median and 95\% credible interval of prevalence trends estimated from the original EPP model without using auxiliary data; the black curves show the posterior median and 95\% credible interval of prevalence trends estimated from EPP augmented by auxiliary data; the black dots show the observed data.}}
\end{figure}